\documentclass[]{elsart}

\usepackage{graphicx}% Include figure files
\usepackage{dcolumn}% Align table columns on decimal point
\usepackage{amsmath}
\usepackage[]{fontenc}
\usepackage{longtable}

\newcommand{\cer}{\v{C}erenkov~}
\newcommand{\acabs}{$D^0 \rightarrow \pi^-\mu^+\nu$~}
\newcommand{\acabf}{$D^0 \rightarrow K^-\mu^+\nu$\ }
\newcommand{\cabs}{$\pi^-\mu^+\nu$~}
\newcommand{\cabf}{$K^-\mu^+\nu$\ }
\newcommand{\cabspi}{$\pi^-\mu^+\nu~ \pi_s^+$~}
\newcommand{\cabfpi}{$K^-\mu^+\nu~ \pi_s^+$\ }
\newcommand{\rom}{$\rho^- \mu^+ \nu$ }
\newcommand{\arom}{$D^0 \rightarrow \rho^- \mu^+ \nu$ }
\newcommand{\kstm}{$K \pi \mu^+ \nu$\ }
\newcommand{\akstm}{$D^0 \rightarrow K \pi \mu^+ \nu$\ }
\newcommand{\kpi}{$D^0 \rightarrow K^-\pi^+$\ }
\newcommand{\ktwopi}{$D^0 \rightarrow K^- \pi^+ \pi^+$\ }
\newcommand{\kpipizero}{$K^-\pi^+\pi^0$}
\newcommand{\kpitwopizero}{$K^-\pi^+\pi^0\pi^0$}
\newcommand{\dmd}{$D^{*+}-D^0$\ }

\newcommand{\relbr}{$\Gamma(\pi^-\mu^+\nu)/\Gamma(K^-\mu^+\nu)$\ }
\begin{document}

\begin{frontmatter}
\title{Measurement of the branching ratio of the decay
 $D^0\rightarrow \pi^-\mu^+\nu$ relative to $D^0\rightarrow K^-\mu^+\nu$}

The FOCUS Collaboration\footnote{See http://www-focus.fnal.gov/authors.html for 
additional author information.}
\author[ucd]{J.~M.~Link},
\author[ucd]{P.~M.~Yager},
\author[cbpf]{J.~C.~Anjos},
\author[cbpf]{I.~Bediaga},
\author[cbpf]{C.~G\"obel},
\author[cbpf]{A.~A.~Machado},
\author[cbpf]{J.~Magnin},
\author[cbpf]{A.~Massafferri},
\author[cbpf]{J.~M.~de~Miranda},
\author[cbpf]{I.~M.~Pepe},
% ...

\address[ucd]{University of California, Davis, CA 95616}
\address[cbpf]{Centro Brasileiro de Pesquisas F\'{\i}sicas, Rio de Janeiro, RJ, Brasil}
\address[cinv]{CINVESTAV, 07000 M\'exico City, DF, Mexico}
\nopagebreak
\address[cu]{University of Colorado, Boulder, CO 80309}
\address[fnal]{Fermi National Accelerator Laboratory, Batavia, IL 60510}
\address[fras]{Laboratori Nazionali di Frascati dell'INFN, Frascati, Italy I-00044}
\address[ugj]{University of Guanajuato, 37150 Leon, Guanajuato, Mexico}
\address[ui]{University of Illinois, Urbana-Champaign, IL 61801}
\address[iu]{Indiana University, Bloomington, IN 47405}
\address[korea]{Korea University, Seoul, Korea 136-701}
\address[kp]{Kyungpook National University, Taegu, Korea 702-701}
\address[milan]{INFN and University of Milano, Milano, Italy}
\address[nc]{University of North Carolina, Asheville, NC 28804}
\address[pavia]{Dipartimento di Fisica Nucleare e Teorica and INFN, Pavia, Italy}
\address[pr]{University of Puerto Rico, Mayaguez, PR 00681}
\address[sc]{University of South Carolina, Columbia, SC 29208}
\address[ut]{University of Tennessee, Knoxville, TN 37996}
\address[vu]{Vanderbilt University, Nashville, TN 37235}
\address[wisc]{University of Wisconsin, Madison, WI 53706}

\date{\today}

\begin{abstract}
We present a new measurement of the branching ratio of
the Cabibbo suppressed decay $D^0\rightarrow \pi^-\mu^+\nu$ 
relative to the Cabibbo favored decay $D^0\rightarrow K^-\mu^+\nu$ and
an improved measurement of the
ratio $\left |\frac{f_+^{\pi}(0)}{f_+^{K}(0)} \right|$.
Our results are $0.074 \pm 0.008 \pm0.007$ for the branching ratio and 
$0.85 \pm 0.04 \pm 0.04 \pm 0.01$ for the form factor ratio, respectively. 
\end{abstract}

\end{frontmatter}

\section{{\normalsize Introduction}\normalsize }

Semileptonic decays provide the advantage of having factorizable weak currents in the Hamiltonian 
which allows for a clean theoretical description. The hadronic current can be described in terms of two form factors,
$f_+(q^2)$ and $f_-(q^2)$ which are functions only of the lepton-neutrino invariant mass squared, $q^2$. Assuming a pole
dominance parameterization of the form factors, we present a parametric analysis of the 
pseudoscalar semileptonic decays \acabs and \acabf from the FOCUS experiment. 

This paper concentrates on the
relative branching ratio and the form factor ratio of the Cabibbo suppressed decay relative
to the Cabibbo favored mode. Since the efficiency tends to have a non-negligible $q^2$ dependence (see
Figure~\ref{fg:effqsq}), we allow the pole masses and the ratio $f_-(0)/f_+(0)$ to vary freely
in the fit. The results and description of the detailed analysis of the
pole masses and $f_-(0)/f_+(0)$ are included in another paper~\cite{JimDorisLore} along with a
non-parametric analysis of the high statistics decay \acabf.

We report a new measurement for the branching ratio \relbr in agreement with recent results 
from the CLEO collaboration~\cite{Huang:2004fr,Gao:2004bw}. These results indicate a lower value 
for this branching ratio than the one reported in the PDG~\cite{Eidelman:2004wy}. 
We also report a new measurement of  the form factor ratio
$\left| \frac{f_+^{\pi}(0)}{f_+^{K}(0)} \right|$ with greatly improved errors with respect
to existing measurements and compare it to recent theoretical predictions from 
an unquenched Lattice QCD calculation~\cite{Okamoto:2003ur,Aubin:2004ej} .

\section{{\normalsize Data Selection}\normalsize }

This analysis is based on data collected by the FOCUS experiment during the 1996--97 
fixed target run at Fermilab. FOCUS is a photoproduction experiment which collected a large
sample of charm decays produced in the interactions of a photon
beam~\cite{Frabetti:1992bn} with an average energy of $\sim$~180 GeV on a BeO segmented target. 
The FOCUS spectrometer~\cite{Frabetti:1990au,Link:2002zg,Link:2002ev,Link:2001pg} is equipped
with a 16 plane silicon strip vertex detector; 4 planes are interleaved with the targets 
and 12 planes are located downstream of the target area. Momentum
analysis is accomplished by two magnets with opposite polarities and 5 multiwire proportional chambers.
Three multi-cell threshold \cer counters provide charged particle identification. A muon counter
located at the end of the spectrometer is responsible for muon identification.

We reconstruct the semileptonic decays 
\acabs and \acabf requiring a $D^{*+}$-tag where the $D^{*+}$ is reconstructed in the $D^0\pi^+$
final state.\footnote{Throughout this paper  charge conjugate modes are implied.}
Whenever possible we apply identical selection criteria to both decay modes
to reduce systematic effects. As the decay \acabs has 
more background and less statistics, the selection cuts have been optimized for this mode.
The signal and normalization samples are selected requiring two 
opposite charged tracks
form a good vertex with a confidence level greater than $1\%$. 
One of the two tracks from the $D^0$ decay  vertex must 
be identified as a muon from the inner muon detector with a confidence level greater than $1\%$
and must have momentum greater than $10$ GeV/$c$. To suppress pion and kaon in-flight decays,
this track is required to have a consistent momentum when measured in the first and second magnets. 
The other track must satisfy a \cer requirement based on the value of the negative log-likelihood
$W$ for a given hypothesis: in the \acabs mode, the pion must be
favored with respect to the kaon hypothesis by at least 3 units of likelihoods ($W(K)-W(\pi)>3$); 
in the case of \cabf the kaon must be favored over the pion hypothesis by 3 units 
of likelihoods ($W(\pi)-W(K)>3$). To reduce non-charm background, the candidate hadron must have a momentum greater than
$14$ GeV/$c$. The primary vertex is found after excluding the candidate tracks from
the $D^0$ decay vertex; the remaining tracks are used to form candidate vertices. Of these vertices
we choose the one with highest multiplicity and we break ambiguities by picking the most upstream vertex
as the primary vertex. This vertex is required to be
isolated from other tracks in the silicon strip vertex detector by requiring that the confidence level 
of any another track not used in the determination of the primary or decay vertex be less than $1\%$. 
For each hadron-lepton combination that satisfies the above requirements,
another track coming from the primary vertex  must be found
as the candidate ``soft" pion from the $D^{*+}\rightarrow D^0 \pi_s^+$. The  $\pi_s^+$ candidate must not have 
the pion hypothesis strongly disfavored over all  other particle hypotheses from the \cer system
($\mathrm{min}\{W(e),W(K),W(p) \}-W(\pi)>-6$).
It must also have a momentum greater than $2.5$ GeV/$c$.
To suppress backgrounds from decays where a final state particle is lost (usually $\pi^0$),
such as \kpipizero, \kpitwopizero, \rom and \kstm\footnote{With the notation \kstm we 
refer to the sum of the decays to $K^0 \pi^- \mu^+ \nu$ and $K^- \pi^0 \mu^+ \nu$ from $D^0$ 
 or to the sum of the decays to $K^- \pi^+ \mu^+ \nu$ and $K^0 \pi^0 \mu^+ \nu$ from $D^+$.}, we place
a lower cut on the hadron-lepton invariant mass (visible mass) of $1.0$ GeV/$c^2$. 
Contamination from \kpi is eliminated by requiring the visible mass to be less than $1.7$ GeV/$c^2$. 

Since the neutrino is not reconstructed, the resultant smearing effects on the resolution 
play an important factor in this analysis. Rather than using the standard neutrino 
closure resulting in a two-fold ambiguity 
on the $D^0$ momentum, we take advantage of the $D^{*+}$-tag by boosting the final state particles in the
hadron-lepton center of mass frame. By constraining the \cabf (\cabs) mass to the $D^0$ mass and the
\cabfpi (\cabspi) mass to the $D^{*+}$ mass, we are able to determine the angle between the 
neutrino\footnote{The neutrino and the $D^0$ directions are the same in this reference frame.} and
the $\pi_s^+$ direction. We then sample the azimuthal angle and choose the one that gives the
direction of the $D^0$  most consistent with pointing to the primary vertex.

\section{{\normalsize Analysis}\normalsize }

\begin{figure}
\centerline{\includegraphics[width=9cm]
{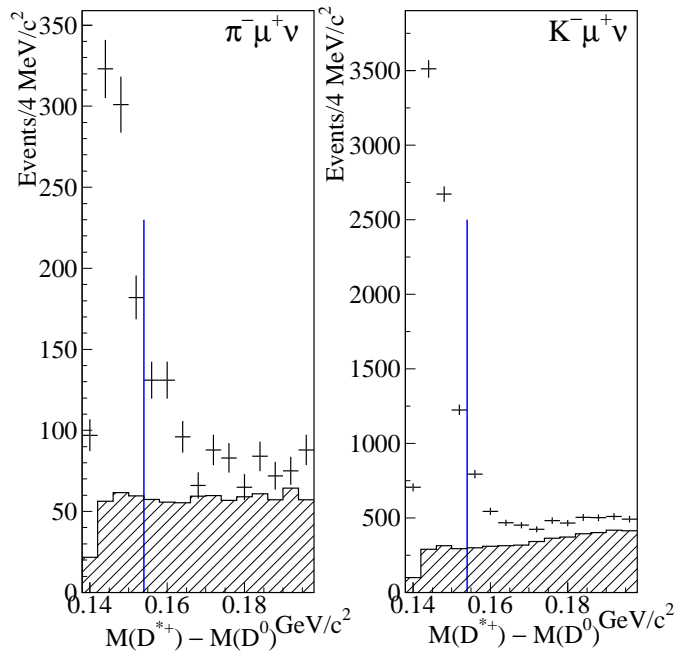}}
\caption[2-dimensional fit projections.]{\dmd mass difference distributions for \acabs (left) and
\acabf (right). The amount of non-peaking background is found from a fit to $q^2$ 
and \dmd mass difference distributions. The vertical line indicates where the cut is placed.}
\label{fg:massdiff}
\end{figure}

The fit to the data is designed to constrain the background in the \cabs sample
and to supply  information about the pole mass and form factors. To accomplish these goals 
we perform fits on two-dimensional distributions where the free parameters are the signal and background 
yields. All the fits are binned maximum likelihood fit where the likelihood is defined as:

\begin{equation}
\mathcal {L}=\prod_{ij}\frac{f_{ij}^{n_{ij}}e^{-f_{ij}}}{n_{ij}!}
\end{equation}

where $f_{ij}$ ($n_{ij}$) is the number of expected (observed) events in the bin $ij$.
First, a fit of $q^2$ and \dmd mass difference is
performed to establish the amount of non-peaking background (Fig.~\ref{fg:massdiff}).\footnote{We define 
``peaking background" in the \cabs sample as the sum of the background contributions from \acabf, \kstm
and \rom, while in the \cabf sample the peaking background is given only by $K^- \pi^0 \mu^+ \nu$.} 
We next place a mass cut on the
\dmd mass difference of less than $0.154$ GeV/$c^2$ to reduce the background and to obtain more reliable 
results for parameters such
as pole masses and form factors. A fit is then made to the two-dimensional distribution $q^2$ vs. $\cos{\theta_\ell}$ 
(where  $\cos{\theta_\ell}$ is defined as the cosine of the angle between the neutrino direction and the
$D^0$ direction in the rest frame of the lepton-neutrino system). The fit is first performed on the \cabf sample  
and the results from this fit are used to set the background from \cabf and \kstm in the \cabs sample. 

In the fit to the \cabf distribution we make use of the recent vector to pseudoscalar branching ratio measurement 
$\Gamma(D^+ \rightarrow K\pi\mu^+\nu)/\Gamma(D^+ \rightarrow \overline{K}\,\!^0 \mu^+ \nu)=0.63 \pm 0.05$~\cite{Link:2004gp}
in the form of a penalty term added to the log-likelihood as shown in Eq.~\ref{eq:like1}:

\begin{equation}
F_{K\mu\nu} = -2~\log{\mathcal{L}_{K\mu\nu}}+ \frac{\left(\frac{Y_{K\pi\mu^+\nu}}{Y_{K^-\mu^+\nu}}
\frac{\epsilon(K^-\mu^+\nu)}{\epsilon(K\pi\mu^+\nu)}-0.63\right)^2}{(0.05)^2}
\label{eq:like1}
\end{equation}

where we assume isospin invariance to relate $D^+$ and $D^0$ decays. 
The likelihood $\mathcal{L}$ is constructed using the expected number of events is each $ij$ bin
of the two-dimensional distribution given by:

\begin{equation}
f_{K^-\mu^+\nu}^{ij}=~Y_{K^-\mu^+\nu}~S_{K^-\mu^+\nu}^{ij}
+Y_{(c\bar c)}~S_{(c\bar c)}^{ij}+Y_{K^-\pi^0\mu^+\nu}~S_{K^-\pi^0\mu^+\nu}^{ij}
\label{eq:cont1}
\end{equation}

where in Eqs.~\ref{eq:like1} and~\ref{eq:cont1} the fit parameters $Y_{\alpha}$ are the fitted yields, $S_{\alpha}$ are the  normalized shapes 
obtained from Monte Carlo and $\epsilon$ the reconstruction efficiency. We define the  $c\bar c$ component as the background 
obtained from a high statistics charm-charmbar Monte Carlo sample after removing the modes handled specifically in Eq.~\ref{eq:cont1} (and~\ref{eq:cont2}).

In a similar way we fit the \cabs distribution. We use the branching ratio 
$\Gamma(D^0 \rightarrow \rho^- \mu^+\nu)/\Gamma(D^0 \rightarrow K\pi\mu^+\nu)=0.086 \pm 0.010$\footnote{We estimated the branching ratio of \arom 
relative to \akstm using the weighted average of a recent result from the 
CLEO-c collaboration~\cite{Gao:2004bw}, the PDG values~\cite{Eidelman:2004wy} and a preliminary 
result from FOCUS~\cite{LuiggiDPF04} where we correct for isospin when necessary.}
to constrain the background in the fit as shown in Eq.~\ref{eq:like2}:

\begin{equation}
F_{\pi\mu\nu} = -2~\log{\mathcal{L}_{\pi\mu\nu}}+\frac{\left(\frac{Y_{\rho^- \mu^+ \nu}}{Y_{K\pi\mu^+\nu}}
\frac{\epsilon(K\pi\mu^+\nu)}{\epsilon(\rho^-\mu^+\nu)}-0.086\right)^2}{(0.010)^2}.
\label{eq:like2}
\end{equation}

The expected number of events in each two-dimensional bin used to construct the likelihood is:

\begin{align}
\nonumber
f_{\pi^-\mu^+\nu}^{ij}&~=~Y_{\pi^-\mu^+\nu}~S_{\pi^-\mu^+\nu}^{ij}+Y_{(c\bar c)}~S_{(c\bar c)}^{ij}+Y_{\rho^-\mu^+\nu}~S_{\rho^-\mu^+\nu}^{ij} +\\
\nonumber
& +Y^0_{K^-\mu^+\nu}~\frac{\epsilon((K^- \rightarrow \pi^-)\mu^+ \nu)}{\epsilon(K^-\mu^+\nu)}~S_{(K^- \rightarrow \pi^-)\mu^+ \nu}^{ij} + \\
\nonumber
& +Y^0_{K^- \pi^0\mu^+\nu}~
\frac{\epsilon((K^- \rightarrow \pi^-)\pi^0 \mu^+\nu)}{\epsilon(K^- \pi^0 \mu^+ \nu)}~S_{(K^- \rightarrow \pi^-)\pi^0 \mu^+\nu}^{ij} + \\
&+2~Y^0_{K^- \pi^0 \mu^+\nu}~
\frac{\epsilon(K^0 \pi^- \mu^+\nu)}{\epsilon(K^- \pi^0 \mu^+ \nu)}~S_{K^0 \pi^- \mu^+\nu}^{ij}. 
\label{eq:cont2}
\end{align}

where $Y^0_{K^-\mu^+\nu}$ and $Y^0_{K^- \pi^0 \mu^+\nu}$ in Eq.~\ref{eq:cont2} are 
fixed to the results obtained from the fit to the \cabf data (Eq.~\ref{eq:cont1}). The symbol 
$(X \rightarrow Y)$ means that a hadron $X$ is misidentified as $Y$.

To measure pole masses and the form factor ratio $\eta \equiv f_-^K(0)/f_+^K(0)$ we apply an event-by-event weighting procedure~\cite{Frabetti:1995xq}. 
This is achieved by re-weighting each Monte Carlo 
event according to the ratio of the probability that the event was generated with a pole mass $M_\mathrm{pole}'$
and a form factor ratio $\eta'$ relative to the probability that the event was generated with the default 
values $M_{D_s^*}$ ($M_{D^*}$ for $\pi\mu\nu$~) and $\eta^0$.~\footnote{The default values 
for the parameter $\eta$ are: $\eta^0=-0.724$ for \acabf and $\eta^0=-0.856$ for \acabs.} 
The relative efficiencies of the decays \acabs and \acabf are defined as the ratio of the reconstructed and
generated Monte Carlo events. At each fit iteration these efficiencies change as a function of the pole 
masses and $\eta$ values. 

The weight $W_i$ for an event with $q^2=q^2_i$ is given by the equation:

\begin{equation}
W_i=\frac{I(M_\mathrm{pole}',\eta ';q_i^2)}{I(M_{D_s^*},\eta^0;q_i^2)}\frac{N(M_{D_s^*},\eta^0)}{N(M_\mathrm{pole}',\eta ')}
%\label{eq:weight}
\end{equation}

where the intensity is:

\begin{equation}
I(M_\mathrm{pole},\eta;q^2) \propto f_+^2(M_\mathrm{pole};q^2)~g(\eta)
\end{equation}

and the normalization is determined by:

\begin{equation}
N(M_\mathrm{pole},\eta)=\sum_{i=1}^\mathrm{Ngen} f_+^2(M_\mathrm{pole};q_i^2)~g(\eta).
\end{equation}

The form factor $f_+(M_\mathrm{pole};q^2)$ is assumed to have the following $q^2$ dependence:

\begin{equation}
f_+(M_\mathrm{pole};q^2)= \frac{f_+(0)}{1-\frac{q^2}{M^2_\mathrm{pole}}}
\end{equation}

and $g(\eta)$ can be written in terms of three kinematic coefficients $\mathcal{A}$, $\mathcal{B}$ 
and $\mathcal{C}$:\footnote{The kinematic dependence is shown in detail for kaon
semileptonic decays in reference~\cite{Eidelman:2004wy} on page 618.}

\begin{equation}
g(\eta)= {\mathcal{A}+\mathcal{B}~\eta+ \mathcal{C}~\eta^2}.
\end{equation}

\begin{figure}
\centerline{\includegraphics[width=10cm]
{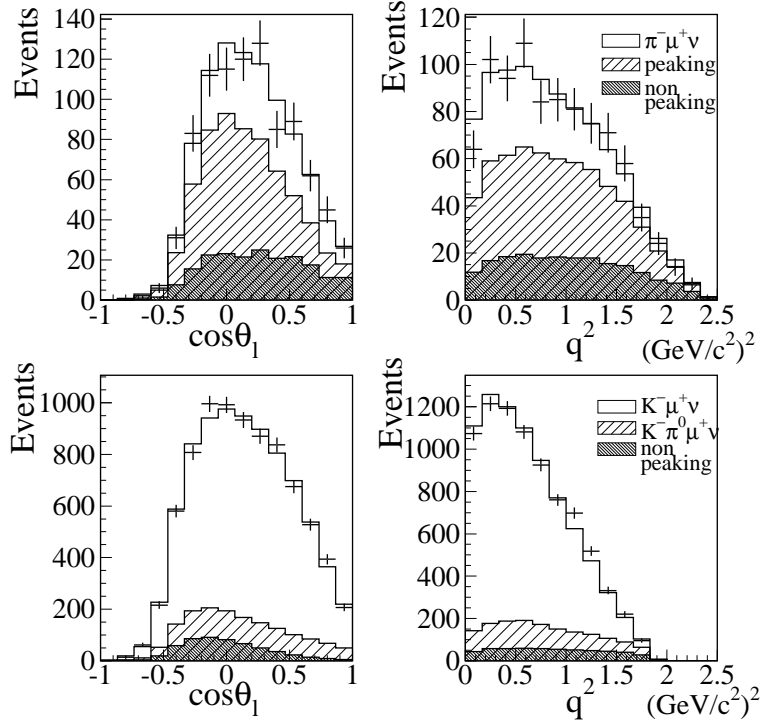}}
\caption[2-dimensional fit projections.]{Fit projections for \cabs and 
$D^0 \rightarrow K^-\mu^+\nu$. The fit
is performed on a two-dimensional distribution of $q^2$ and $\cos{\theta_\ell}$. 
In the $D^0 \rightarrow \pi^-\mu^+\nu$,
the peaking background contribution is defined as the sum of the contributions from
$D^0 \rightarrow K^-\mu^+\nu$, \rom and $K\pi \mu^+ \nu$.}
\label{fg:fit}
\end{figure}

From the  fit to the \cabs ($K^-\mu^+\nu$) distributions (Fig.~\ref{fg:fit}) we find $288 \pm 29$ 
\cabs ($6574  \pm 92$ $K^-\mu^+\nu$) events. Correcting for the relative Monte Carlo 
efficiency we find the branching ratio for the Cabibbo suppressed decay \acabs relative to the
Cabibbo allowed decay \acabf to be:

\begin{equation}
\frac{\Gamma (D^{0}\! \rightarrow \! \pi^{-}\mu^+\nu)}{\Gamma (D^{0}\! \rightarrow \! K^{-}
\mu^{+}\nu)}=0.074\pm 0.008~(\mathrm{stat.}).
\end{equation}

From the same fits we find $M_{\pi}=1.91^{+0.30}_{-0.15}$ and 
$M_{K}=1.93^{+0.05}_{-0.04}$ for the \cabs and the \cabf pole masses respectively. 
We also measure the ratio $f_-^{K}(0)/f_+^{K}(0)=-1.7^{+1.6}_{-1.4}$. A
detailed description of the pole mass results has been included in Ref.~\cite{JimDorisLore}. 

\begin{figure}
\centerline{\includegraphics[width=10cm]
{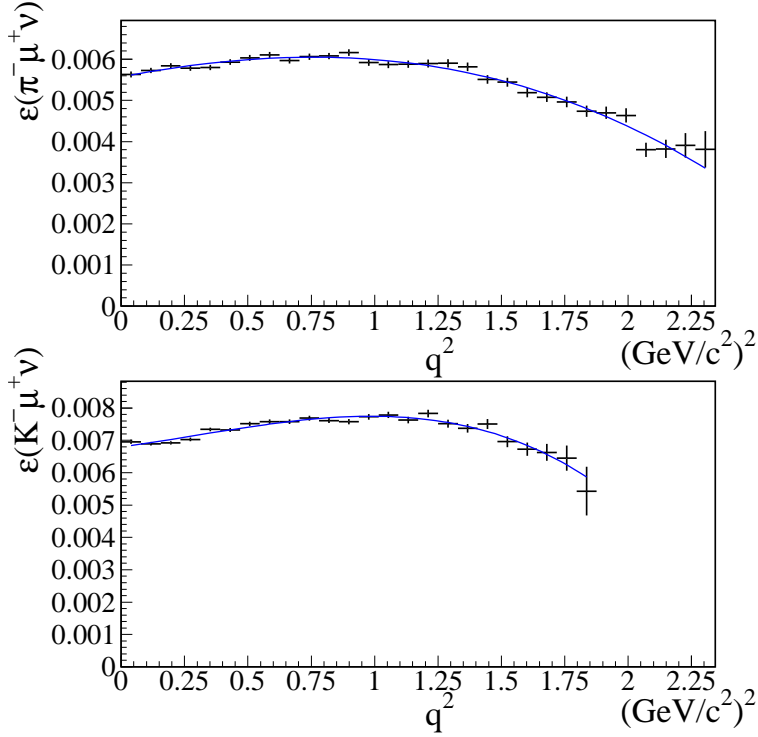}}
\caption[2-dimensional fit projections.]{Reconstruction efficiency as a function of the $q^2$ for
\cabs (top) and \cabf (bottom). }
\label{fg:effqsq}
\end{figure}

Using the yields from the fit it is possible to obtain the ratio of the form factors 
$f_+^{\pi}(0)/f_+^{K}(0)$. In
order to do this we compute a numerical integration of the differential decay rate modulated by 
the reconstruction efficiency as a function of the $q^2$~\cite{Frabetti:1996jj}. This efficiency 
is found by sampling the
$q^2$ Monte Carlo distribution and dividing the reconstructed events by the generated events in each bin. 
The resultant distribution is then fit to a third degree polynomial (Fig.~\ref{fg:effqsq})
which is used in the computation of the integral.
We quote the result: 

\begin{equation}
\left|\frac{V_{cd}}{V_{cs}} \right|^2 \left|\frac{f_+^{\pi}(0)}{f_+^{K}(0)} \right|^2=0.037 \pm 0.004~(\mathrm{stat.}).
\label{eq:ffratio1}
\end{equation}

Applying the unitarity constraints on the CKM matrix elements~\cite{Eidelman:2004wy} we use the 
value $|\frac{V_{cd}}{V_{cs}}|^2=0.051\pm0.001$ in Eq.~\ref{eq:ffratio1} and measure 
the ratio $f_+^{\pi}(0)/f_+^{K}(0)$ to be: 

\begin{equation}
\left|\frac{f_+^{\pi}(0)}{f_+^{K}(0)} \right|=0.85\pm 0.04~(\mathrm{stat.}).
\end{equation}

\section{{\normalsize Systematic Studies}\normalsize }

Several studies have been performed to search for possible systematic uncertainties. 
The fitting procedure was tested on a Monte Carlo set whose size is roughly 20 times the
FOCUS data set and we verified that the fit returned the input values used in our simulation. 

We checked for possible biases as well as the accuracy of our statistical error by performing a fit
on fluctuated data distributions multiple times and comparing the mean and width of the distribution
of the fit results to our measurement. We found that we have to add a 0.005 contribution
to the systematic error to compensate for \cabf and \kstm contributions that were not allowed to float
in the \cabs fit. We also performed an analogous study using the fit function as the parent
distribution to establish how well our fit function described the data. We compared the likelihood
obtained from our measurement to the distribution of the likelihoods from the fluctuated fit
function. We found good agreement indicating that our fit function 
well represents the data.

We investigated the stability of our results by changing a variety of selection
criteria: the significance of separation between the primary and secondary vertex, muon identification,
track momenta, visible mass cut, and \cer identification. We found no significant change
in our results and assign a systematic uncertainty of 0.003 on the branching ratio
due to cut variations. This number is found by computing the variance of this set of results.

We further investigated fit variations by using a different approach in
which we fit the $q^2$ and \dmd mass difference. Rather than fitting the \cabf distribution first, this
fit was performed simultaneously on the \cabs and \cabf samples. The results are nearly 
identical to the results obtained from the fit to $q^2$ and $\cos{\theta_\ell}$. 
Other fit variations include changing the bin size. By computing the variance of these 
a priori likely results, we assigned a systematic uncertainty of 0.004 from 
fit variations.

Since the Monte Carlo is used to determine the amount of \cabf background in the \cabs sample,
we are sensitive to the simulated misidentification rate. We used the high statistics modes \kpi 
and \ktwopi where no \cer requirement was applied to measure the   
$K\rightarrow\pi$ misidentification rate and we used the statistical error on the combined sample 
(after applying the same \cer requirement used to select the \acabs events) to assign a systematic uncertainty. 
We varied the misidentification rate so obtained by $\pm 1\sigma$, and we
find a contribution of 0.002 to the total systematic error. 

The contributions to the systematic error on the branching ratio and the corresponding contributions
to the error on the form factor ratio $f_+^{\pi}(0)/f_+^{K}(0)$ are listed in Table~\ref{tb:system}.

\begin{table}[htbp]
\begin{center}
\begin{tabular}{|l|l|c|} \hline

      & BR      &  $f^{\pi}_{+}(0)/f^{K}_{+}(0)$\\ \hline \hline

Fluctuated data distribution         & 0.005  & $0.029$  \\ \hline
Cut variations                       & 0.003  & $0.017$  \\ \hline
Fit variations                       & 0.004  & $0.023$  \\ \hline
\cer misidentification               & 0.002  & $0.011$  \\ \hline
%$\epsilon(q^2)$ fit variation       &        & $0.010$  \\ \hline\hline
Total                                & 0.007  & $0.042$   \\ \hline\hline
 
\end{tabular}
\caption[Systematic uncertainties.]{Sources of systematic errors and relative uncertainties.
The contributions to the error on the ratio $f^{\pi}_{+}(0)/f^{K}_{+}(0)$ are found by propagating
the corresponding errors on the branching ratio.}
\label{tb:system}
\end{center}
\end{table}

In the measurement of the form factor ratio $f^{\pi}_{+}(0)/f^{K}_{+}(0)$ we also added variations on the fit 
to the reconstruction efficiency as a function of the $q^2$ used in the numerical 
integration. We varied the bin size and fit functions. 
We find a contribution to the systematic error of 0.010  which is added in quadrature to the errors
propagated from the branching ratio measurement.  

\section{Summary and Conclusions}

We quote the final results as:
\begin{equation}
\frac{\Gamma (D^{0}\! \rightarrow \! \pi^{-}\mu^+\nu)}{\Gamma (D^{0}\! \rightarrow \! K^{-}
\mu^{+}\nu)}=0.074\pm 0.008~(\mathrm{stat.}) \pm 0.007~(\mathrm{sys.})
\end{equation}

and 

\begin{equation}
\left|\frac{V_{cd}}{V_{cs}}\right|^2 \left|\frac{f_+^{\pi}(0)}{f_+^{K}(0)}\right|^2=0.037 \pm 0.004~(\mathrm{stat.}) \pm 0.004~(\mathrm{sys.}).
\end{equation}

Using $|\frac{V_{cd}}{V_{cs}}|^2=0.051\pm0.001$ from unitarity constraints, we find the form factor ratio to be:

\begin{equation}
\left|\frac{f_+^{\pi}(0)}{f_+^{K}(0)} \right|=0.85\pm 0.04~(\mathrm{stat.})\pm 0.04~(\mathrm{sys.})\pm 0.01~(\mathrm{CKM})
\end{equation}

where the last error (CKM) corresponds to the uncertainty on the ratio $|V_{cd}/V_{cs}|$. We compare our results to
the measurement reported by the CLEO collaboration in Ref.~\cite{Huang:2004fr} where they report the branching ratio
of $D^0 \rightarrow \pi^- e^+\nu$ relative to $D^0 \rightarrow K^- e^+\nu$ to be $0.082 \pm 0.006 \pm 0.005$ and 
the form factor ratio $\left|\frac{f_+^{\pi}(0)}{f_+^{K}(0)}\right|=0.86 \pm 0.07^{+0.06}_{-0.04} \pm 0.01$. We also
compare our branching ratio result to the recent measurement from absolute branching ratios
for $D^0 \rightarrow \pi^- e^+\nu$ and $D^0 \rightarrow K^- e^+\nu$ from CLEO-c~\cite{Gao:2004bw} where they report
a relative branching ratio of $0.070 \pm 0.007 \pm 0.003$. Our results are consistent with both of these new measurements.
Further, we report an improved measurement of $\left|\frac{f_+^{\pi}(0)}{f_+^{K}(0)} \right|$ in good agreement 
with SU(3) breaking expected in recent lattice QCD calculations where they
quote a form factor ratio value of $0.85 \pm 0.05$~\cite{Okamoto:2003ur} and $0.86 \pm 0.05 \pm 0.11$~\cite{Aubin:2004ej}.

\section{Acknowledgments}
We wish to acknowledge the assistance of the staffs of Fermi National
Accelerator Laboratory, the INFN of Italy, and the physics departments of the
collaborating institutions. This research was supported in part by the U.~S.
National Science Foundation, the U.~S. Department of Energy, the Italian
Istituto Nazionale di Fisica Nucleare and Ministero dell'Universit\`a e della
Ricerca Scientifica e Tecnologica, the Brazilian Conselho Nacional de
Desenvolvimento Cient\'{\i}fico e Tecnol\'ogico, CONACyT-M\'exico, the Korean
Ministry of Education, and the Korean Science and Engineering Foundation.

\bibliographystyle{myapsrev}
\bibliography{main}

\end{document}